\documentclass[a4paper]{article}

\usepackage{INTERSPEECH2021}

\title{Dual-Path Filter Network: Speaker-Aware Modeling for Speech Separation}
\name{Fan-Lin Wang$^{1,2}$, Yu-Huai Peng$^1$, Hung-Shin Lee$^{1,2}$, and Hsin-Min Wang$^1$}
\address{
$^1$Institute of Information Science, Academia Sinica, Taiwan\\
$^2$Department of Electrical Engineering, National Taiwan University, Taiwan}
\email{faliwang1999@gmail.com}

\begin{document}

\maketitle

\begin{abstract}

Speech separation has been extensively studied to deal with the cocktail party problem in recent years. All related approaches can be divided into two categories: time-frequency domain methods and time domain methods. In addition, some methods try to generate speaker vectors to support source separation. In this study, we propose a new model called dual-path filter network (DPFN). Our model focuses on the post-processing of speech separation to improve speech separation performance. DPFN is composed of two parts: the speaker module and the separation module. First, the speaker module infers the identities of the speakers. Then, the separation module uses the speakers' information to extract the voices of individual speakers from the mixture. DPFN constructed based on DPRNN-TasNet is not only superior to DPRNN-TasNet, but also avoids the problem of permutation-invariant training (PIT).
\end{abstract}

\section{Introduction}

For decades, researchers have been studying the cocktail party problem \cite{Haykin2005}, which refers to the problem of perceiving speech in a noisy social environment. The human auditory system has the brilliant ability to extract a target source from a complex mixture. This is what the automatic system is trying to learn. For many downstream tasks of speech processing, such as speaker diarization \cite{Neumann2019} and automatic speech recognition \cite{Raj2020a}, speech separation is a necessary pre-processing.

Multi-speaker monaural speech separation aims to isolate the voices of individual speakers from a recording of completely overlapping voices. All methods in recent years can be briefly grouped into two categories: time-frequency (TF or spectrogram) domain methods  \cite{Hershey2016,Huang2014,Kolbk2017,Wang2018c,Chen2017,Luo2018a} and time-domain methods \cite{Venkataramani2019,Zhang2020,Luo,Luo2018,Nachmani2020,Chen2020}. The former is to use short-time Fourier transform (STFT) to convert the time-domain mixture into the time-frequency domain for separation. Usually, deep neural network (DNN)-based methods, such as deep clustering (DPCL) \cite{Hershey2016,Isik2016}, permutation-invariant training (PIT) \cite{Yu2017}, and Deep CASA \cite{Liu2019a}, are used to estimate the ideal binary mask (IBM), ideal ratio mask (IRM), or phase-sensitive mask (PSM). In the final step , the source waveform will be reconstructed by inverse STFT. However, the accurate phase of a clean source is difficult to reconstruct, which imposes an upper limit on the separation performance. Therefore, many researchers tend to use the time-domain approach. 

Time-domain methods use the original waveform as input. The most popular method is based on time-domain audio separation network (TasNet) \cite{Luo2018}. This model is composed of an encoder, a separator, and a decoder. The encoder (Conv1D) converts the waveform into embedding, while the decoder (TransposeConv1D) converts the embedding processed by the separator into a waveform. The main focus of TasNet is the separator that estimates the masks. A lot of work has since been done to improve the separator, such as fully-convolutional TasNet (Conv-TasNet) \cite{Luo}, dual-path recurrent neural network (DPRNN) \cite{Luo2020}, gated DualPathRNN \cite{Nachmani2020}, dual-path transformer network (DPT-Net) \cite{Chen2020}, and SepFormer \cite{Subakan2021}. Among them, the dual-path method is the mainstream, which processes the waveform from the two dimensions of the local path and the global path. Although the performance of the time-domain methods surpasses that of the TF methods, the processing of real-world recordings is still challenging. For instance, most speech separation models can only deal with recordings of a fixed number of speakers.

To solve the limitation, inspired by speech extraction \cite{Delcroix2018a,Xu2020,Xu2018a,Wang2019a}, researchers began to use speaker information to support speech separation. As long as we extract one target speaker at a time, the separation model is not constrained by the fixed number of speakers. In this way, the entire model is composed of two parts: a speaker module and a separation module. The speaker module takes the mixed or initially separated waveform as input and provides the speaker identity to the separation module. There are many ways to incorporate the speaker identity into the separation module. For instance, the speaker vector is affine transformed in the separation module in WaveSplit \cite{Zeghidour,Perez2018}, concatenated with the frame-wise feature vectors of the mixture in Speaker-Conditional Chain Model (SCCM) \cite{Shi2020}, or simply used to calculate a speaker classification loss in TasTas \cite{Shi2020a}. Either way above can improve the performance of speech separation.

In this paper, we propose a method called dual-path filter network (DPFN). It is a filter-based model, which means that it acts as a voice filter to pass the target source waveform in the mixture. It combines the advantages of the dual-path method and the use of speaker information. In DPFN, the speaker module is newly designed and inspired by the separation model. We believe that it can produce more helpful speaker embeddings for speech separation. Moreover, based on its model structure and characteristics of post-processing, DPFN can be connected to any separation model to improve separation performance.

The contribution of this paper spans the following aspects: (1) We build the first filter-based model that focuses on the post-processing of speech separation. (2) Our model provides a brand new speaker module to produce representative speaker filters. (3) Under the same basic structure, our separation model gives a better performance than DPRNN. (4) We do not need to use permutation-invariant training (PIT) in the training phase.

\section{Related Work}
\subsection{DPRNN}
The dual-path recurrent neural network (DPRNN) \cite{Luo2020} is mainly composed of three parts: the encoder, the separator, and the decoder. First, the encoder (Conv1D) takes the mixture as input and transforms it into the corresponding representation. Then, the representation is input to the separator, which in order consists of a LayerNorm, a 1×1conv, 6 dual-path BiLSTM layers, PReLU, 1×1conv, and a sigmoid operation, to estimate the masks of individual sources. Finally, the decoder (TransposeConv1D) is used to reconstruct the source waveform from the masked encoder features.

The main feature of DPRNN-TasNet is the local and global data chunk formulation in the dual-path BiLSTM module. The model first splits the output of the encoder into chunks with or without overlaps and concatenates them to form a 3-D tensor. The dual-path BiLSTM module will map the 3-D tensor to the 3-D tensor masks of individual speakers. Then, the product of each mask and the original 3-D tensor is converted back to a sequential output by an ‘Overlap-Add’ operation. The BiLSTM module goes through two different dimensions: the chunk size (intra-chunk RNN) and the number of chunks (inter-chunk RNN). Therefore, RNN can see the information around and far from the current time frame. This is the main factor for its excellent performance and the reason why we decided to use DPRNN-TasNet as the basis of our separation module.

\subsection{SCCM}

The speaker-conditional chain model (SCCM) \cite{Shi2020} is named after its chain of two sequential components: speaker inference and speech extraction. The speaker inference module aims to predict the possible speaker identities and the corresponding embedding vectors. Here, the module uses a self-attention-based Transformer and takes the mixture spectrogram (STFT coefficients) as input. Then, the speech extraction module extracts the corresponding source audio from the input recording based on the embedding of each speaker. More specifically, the speaker embedding is concatenated with the frame-wise spectral features. A structure similar to Conv-TasNet is used as the speech extraction module, but one source is extracted at a time. Inspired by SCCM, our model is also composed of speaker inference and speech extraction, but the details of modules are completely different.

\section{DPFN}

\begin{figure}[t]
\begin{center}
\includegraphics[width=0.48\textwidth]{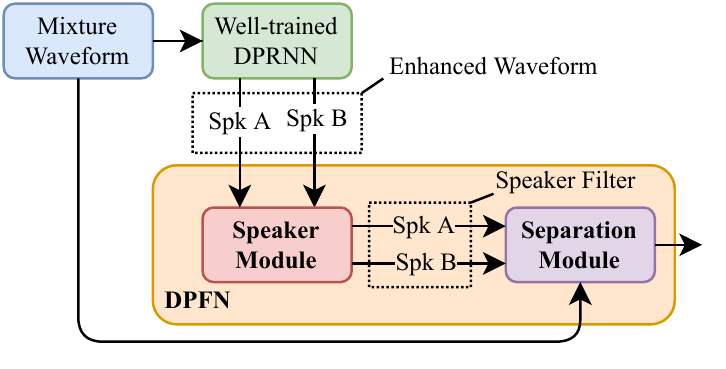}
\end{center}
\vspace{-15pt}
\caption{The structure of DPFN.}
\label{fig:dpfn}
\vspace{-15pt}
\end{figure}

\begin{figure}[t]
\begin{center}
\includegraphics[width=0.27\textwidth]{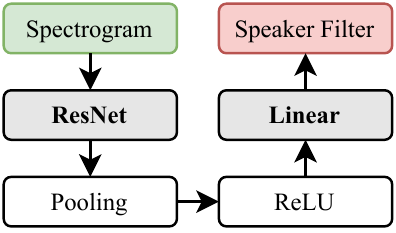}
\end{center}
\vspace{-15pt}
\caption{The structure of the speaker module for the situation where the speakers are unknown.}
\label{fig:spk_module}
\vspace{-0pt}
\end{figure}

\begin{figure}[t]
\begin{center}
\includegraphics[width=0.48\textwidth]{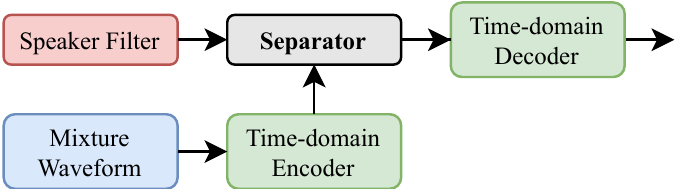}
\end{center}
\vspace{-15pt}
\caption{The structure of the separation module.}
\label{fig:sep_module}
\vspace{-15pt}
\end{figure}

\subsection{Model Design}

The proposed dual-path filter network (DPFN) is a filter-based model. We focus on the post-processing of speech separation and extract the target speech with the speaker filter as the condition. As shown in Figure \ref{fig:dpfn}, we cascade DPFN to a pre-trained separation model to filter a cleaner source waveform. There are two situations: one is that the speakers' identities are known and we have their clean recordings to extract their speaker embeddings; the other is that we only have the mixed waveform and do not know who the speakers are. No matter in which scenario it is used, DPFN is composed of two parts: a speaker module and a separation module. We use DPRNN as the basis of the separation module. In the following, we will introduce the model structure in three parts: the speaker module when the speakers' identities are known, the speaker module when the speaker information is unknown, and the separation module.
\subsubsection{Speaker Module for Known Speakers}
In this scenario, we use a speaker module to obtain the speaker filters. We believe that the most representative and text-independent speaker filter so far is x-vector \cite{Snyder2018c}. Therefore, we use the pre-trained SRE16 x-vector model from Kaldi as the speaker module. We used the recordings in the Wall Street Journal (WSJ0) dataset to obtain the x-vector of each speaker in the dataset. The x-vector is passed through a trainable fully-connected layer to increase flexibility, and then used as the speaker condition by the separation module. 
\subsubsection{Speaker Module for Unknown Speakers}
When only the mixed waveform is given without the speaker information, we first separate the mixture through a pre-trained separation model and input the separated waveforms into our speaker module to obtain speaker filters. Our speaker module is not restricted by the number of speakers. Each time we input a speaker's separated speech into the speaker module, it will produce the corresponding speaker filter. Then, we can get the correct number of speeches by the speaker filters. Inspired by Conv-TasNet, the structure of our speaker module is mainly composed of a Residual Net and a pooling layer, as shown in Figure \ref{fig:spk_module}. First, the input waveform $\boldsymbol{X}$ is transformed into its spectrogram, which contains $\Tilde{T}$ frames and $F$ frequency bins. Then, the spectrogram is sent to the Residual Net. Finally, the output passes through a pooling layer to obtain the speaker filter, which will be used by the separation module.

The Residual Net is composed of $S$ stacks, and each stack contains a 1D convolution layer, $B$ residual blocks, and a nonlinear activation function, as expressed in Eqs. (\ref{eq_conv1d}) to (\ref{eq_relu}). The spectrogram input $\mathbf{X}_0\in \mathbb{R}^{F\times \Tilde{T}}$ is sent to the first stack.
\begin{equation} \label{eq_conv1d}
  \mathbf{Y}_{i,0} = Conv1d(\mathbf{X}_i) \qquad i = 0,...,S-1
\end{equation}
\begin{equation} \label{eq_res}
  \mathbf{Y}_{i,j+1} = ResBlock(\mathbf{Y}_{i,j}) \qquad j = 0,...,B-1
\end{equation}
\begin{equation} \label{eq_relu}
  \mathbf{X}_{i+1} = LeakyReLU(\mathbf{Y}_{i,B})
\end{equation}
In each residual block, the input $\mathbf{Y}_{i,j}$ is processed by a nonlinear activation function, a 1-D convolution, and a normalization operation to get the output $\mathbf{Y}_{i,j+1} \in \mathbb{R}^{\Tilde{R'} \times \Tilde{T}}$ as,
\begin{equation} \label{eq_res_net}
  \mathbf{Y}_{i,j+1} = LayerNorm(Conv1d(LeakyReLU(\mathbf{Y}_{i,j}))).
\end{equation}
Note that every two blocks, a residual path is added to the end of the block. This is why we call it a residual net.

After $S$ stacks, a linear operation is applied to the output $\mathbf{X}_S$ to get $\mathbf{Z} \in \mathbb{R}^{\Tilde{R} \times \Tilde{T}}$.
Then, a pooling layer is used to compute the mean of the last dimension (i.e., time frames) of $\mathbf{Z}$: 
\begin{equation} \label{eq_pool}
  \mathbf{Z'} = Pooling(\mathbf{Z}).
\end{equation}
Finally, a nonlinear activation function and a linear operation are applied to produce the speaker filter $\mathbf{V} \in \mathbb{R}^{D}$, where $D$ is the embedding dimension of the speaker filter:
\begin{equation} \label{eq_pool_relu}
  \mathbf{V} = Linear(ReLU(\mathbf{Z'})).
\end{equation}

\subsubsection{Filter Based Separation Module}
Inspired by DPRNN-TasNet, the separation module is composed of three parts: the encoder, the separator, and the decoder, as shown in Figure \ref{fig:sep_module}. The encoder and the decoder are the same as those in DPRNN-TasNet, so we focus on the separator here. The separator consists of three stages: segmentation, block processing, and overlap-add. Among these stages, we add the speaker condition to block processing. The $L$ DPRNN blocks can be divided into two categories: inter-chunk and intra-chunk. The difference between these two is that the chunks segmented and stacked from the input are sliced in two different directions: chunk size $C$ and number of chunks $N$. In the block processing of our separation module, the chunk $R$ goes through an LSTM and a fully-connected layer: 
\begin{equation} \label{rnn}
  \mathbf{R}'_i = \mathbf{G}(BiLSTM(\mathbf{R}_i))+\mathbf{m} \qquad i = 0,...,L-1
\end{equation}
The fully-connected layer here can make the processed chunk $\mathbf{R}'$ maintain the original dimension of $\mathbf{R}$, where the weight is $\mathbf{G}$ and the bias is $\mathbf{m}$.

Then, we add the speaker filter $\mathbf{V}$ as a condition here. We believe that the speaker filter is a centroid in the embedding space. Therefore, we use FiLM \cite{Perez2018} to add the condition, which can attract the embedding vector to get closer to the speaker's centroid. The method here is to transform the speaker filter $\mathbf{V}$ into the weight $c_1$ and bias $c_2$ by a linear operation (Eq. (\ref{cc})). Then, the chunk multiplies the weight and adds the bias of a specific speaker, and then goes through a nonlinear activation function (Eq. (\ref{prelu})). Finally, we apply normalization to the chunk and add a residual path at the end of the block, which is the same as DPRNN (Eq. (\ref{norm})).
\begin{equation} \label{cc}
  c_1 = Linear(\mathbf{V}), \quad c_2 = Linear(\mathbf{V})
\end{equation}
\begin{equation} \label{prelu}
  \mathbf{R}''_i = PReLU(c_1 \times \mathbf{R}'_i + c_2)
\end{equation}
\begin{equation} \label{norm}
  \mathbf{R}_{i+1} = LayerNorm(\mathbf{R}''_i) + \mathbf{R}_i
\end{equation}

\subsection{Training Targets}
In our model, there are two training criteria: one for knowing speakers' identities and the other for not knowing the speakers' identities. For the former criterion, we trained the separation module and the fully-connected layer, the latter being connected to the fixed pre-trained speaker model. In this situation, we only consider the source reconstruction loss $\mathcal{L}_r$ in terms of the scale-invariant signal-to-distortion ratio (SI-SNR). If we denote the output of the network by $x$, which should ideally be equal to the target source $s$, SI-SNR can be calculated as
\[
    \tilde{x}=\frac{\langle x, s\rangle}{\langle x, x\rangle} x, \quad e=\tilde{x}-s, \quad \text { SI-SNR }=10 * \log _{10} \frac{\langle\tilde{x}, \tilde{x}\rangle}{\langle e, e\rangle}.
\]
Our target is to maximize SI-SNR or minimize the negative SI-SNR. Unlike most work in speech separation, we do not need to perform permutation-invariant training (PIT). This is because we already know the identities of the speakers, and we can compare the original source waveform with the separated waveform.

On the other hand, in the absence of speaker information, we jointly trained the speaker module and the separation module. In addition to the source reconstruction loss, we also tried to use the speaker identity loss $\mathcal{L}_i$, which is a cross-entropy loss. However, regardless of the loss weight, the training results were all overfitting. Therefore, the experiment results reported in this paper are the results of using only the source reconstruction loss $\mathcal{L}_r$. Moreover, we first pre-trained the speaker module with clean speech, and then fine-tuned the model with the separated speech from DPRNN-TasNet. Most importantly, we did not use PIT for training in this case, because we believe in the alignment of the separated speech and the clean speech from DPRNN-TasNet. That is, when we put the target speaker's separated speech into the speaker module and give the speaker filter to the separation module, the separation module should produce the target speaker's speech and vice versa. There are several advantages for not using PIT. First, we can reduce the complexity of loss calculation and make training faster. Second, since the calculation process is simplified, we can add a variety of losses to the training target, such as phonetic loss or content-aware loss, to help improve the performance of speech separation. 

\begin{table}[t]
\caption{Speech separation performance (SI-SNR) on WSJ0-2mix with known speaker information.}
\setlength{\tabcolsep}{10pt}
\vspace{-5pt}
\label{tab:know}
\footnotesize
\centering
\begin{tabular}{lcc}
\toprule
\bf{X-vector} & \bf{SI-SNR (dev)} & \bf{SI-SNR (eval)} \\ 
\midrule\midrule
Target & 16.21 & 15.36 \\
Non-Target & 16.15 & 15.24 \\
Both & 16.86 & 16.13 \\
Baseline (DPRNN) & \bf{17.82} & \bf{17.40} \\
\bottomrule
\end{tabular}
\vspace{-15pt}
\end{table}

\section{Experiments}
\subsection{Datasets}
We evaluated our filter-based network on the two-speaker speech separation problem, although we can easily expand the number of speakers. To extract the target speech, our model only needs the speaker's filter, which can be obtained from the initially separated speech provided by any speech separation model. We used the WSJ0-2mix dataset, which is a benchmark dataset for two-speaker mono speech separation in recent years. WSJ0-2mix contains 30 hours of training data and 10 hours of validation (development) data. The mixtures are generated by randomly selecting the utterances of 101 speakers in the Wall Street Journal (WSJ0) training set si\_tr\_s and mixing them at various signal-to-noise ratios (SNR) uniformly between 0 dB and 5 dB (the SNRs for different pairs of mixed utterances are fixed as in \cite{Hershey2016} for fair comparison). 5 hours of evaluation data are generated in the same way, using utterances from other 16 unseen speakers from si\_dt\_05 and si\_et\_05 in the WSJ0 dataset. In the speech separation domain, all the recordings are downsampled to 8k Hz. In addition, for our speaker module, the magnitude spectrogram computed from STFT with 160 ms window length, 80 ms hop size, and the Hann window is used as the input feature.

\begin{table}[t]
\caption{Comparison of the performance of the model with fine-tuning using the initially separated speech (sep) and the model without fine-tuning (clean).}
\setlength{\tabcolsep}{10pt}
\vspace{-5pt}
\label{tab:compare}
\footnotesize
\centering
\begin{tabular}{l|cc|cc}
\toprule
&\multicolumn{2}{c|}{\bf{dev}} & \multicolumn{2}{c}{\bf{eval}} \\
& \bf{clean} & \bf{sep} & \bf{clean} & \bf{sep} \\
\midrule\midrule
Target & 17.66 & 17.75 & 16.13 & 16.04\\
Non-Target & 17.56 & 17.74 & 15.97 & 15.49 \\
Both & 19.56 & \bf{19.76} & 18.74 & \bf{18.84} \\
\bottomrule
\end{tabular}
\vspace{-5pt}
\end{table}

\subsection{Results}
Our experiments were divided into two parts based on the two scenarios. Because the separation module is based on DPRNN-TasNet and the initially separated speech is provided by the pre-trained DPRNN-TasNet, we used DPRNN-TasNet as our baseline model. Here, the implementation of DPRNN-TasNet was from the Asteroid toolkit \cite{Pariente2020}. We followed the script and trained the DPRNN model to get the separated speech and the baseline performance. In this toolkit, when DPRNN-TasNet performs inference, it calculates pairwise SI-SDR and aligns the separated speech with the corresponding original source that gives the higher SI-SDR. This is how we got the pair of separated speech and original speech to avoid PIT. 

In addition, our filter is speaker-aware. We tried three different speaker-aware approaches. The first one is that we use the target speaker's filter as the condition and the separation module produces the target speaker's speech accordingly. This way is the most popular. The second one is that we use the non-target speaker filter as the condition but make the separation module to produce the target speaker's speech. The reason why we tried this method is that DPRNN-TasNet is a mask-based separation model. We thought that the main information the mask should really learn is all the information in the waveform except the target speaker. The third method is that we concatenate the filters of the two speakers as the condition for the separation module to simultaneously generate two corresponding separated voices.  The dimension of DPRNN was modified accordingly.

\subsubsection{Experiments with Known Speaker Information}
Table \ref{tab:know} shows the results of the first part of experiments. We can see that the performance on the development set is slightly better than that on the evaluation set. The reason may be that the speakers in the development set and the training set overlap, while the speakers in the evaluation set are completely unseen to the separation model. However, the difference is so small that we can still conclude that the model has successfully learned how to separate speech from unseen speakers. On the other hand, among the three different types of filters, the performance of providing target speaker filters and providing non-target speaker filters are comparable. We think the reason is that the function of the mask in the speech that contains only two speakers is relatively simple. No matter which information is provided, the separator can learn how to estimate a fine mask. To learn more about how the mask-based separator learns to produce masks, we should consider speech with more types of interference, such as noise or a larger number of speakers. Among the three methods, the performance of the last method is the best. It is reasonable because in this case, the separation module uses the information of the two speakers at the same time, and it is easier to see the difference between the two speakers. Finally, we can see that the performance of our x-vector-based models is inferior to the performance of the original DPRNN-TasNet. We think that the main reason is that although the x-vector is robust for speaker recognition, it is not necessarily robust for speech separation. A simple fully-connected layer may not be enough to transform the x-vector into an embedding that is more suitable for speech separation.

\subsubsection{Experiments without Speaker Information}


In the second part of experiments, we first studied the effectiveness of fine-tuning the speaker and separation models using the initially separated speech. The results are shown in Table \ref{tab:compare}, where $clean$ means that we only used the clean speech as the input of the speaker module to jointly train the speaker and separation models; while $sep$ means that we first used the clean speech as the input of the speaker module to jointly train the speaker and separation models, and then used the initially separated speech as the input of the speaker module to fine-tune the speaker and separation models. From the table, we can see that fine-tuning the models with the initially separated speech can improve the performance. 

Table \ref{tab:noknow} shows the performance of speech separation without speaker information. We chose the speaker and separation models fine-tuned with the initially separated speech. Among the three filters, the last one stands out. We have also observed the same trend in Table \ref{tab:know}. However, we can see that the overall performance in Table \ref{tab:noknow} is much better than the performance in Table \ref{tab:know}. The results show that our speaker module can successfully generate suitable speaker filters for speech separation. In addition, it is worth noting that our model based on the third filter outperforms the original DPRNN, which proves the effect of our model as a post-processing of DPRNN. Although more parameters in our model may lead to a greater risk of overfitting, our current work did not encounter this problem.

\begin{table}[t]
\caption{Speech separation performance (SI-SNR) on WSJ0-2mix without speaker information.}
\vspace{-5pt}
\label{tab:noknow}
\footnotesize
\centering
\begin{tabular}{lcc}
\toprule
\textbf{Proposed} & \bf{SI-SNR (dev)} & \bf{SI-SNR (eval)} \\ 
\midrule\midrule
Target & 17.75 & 16.04\\
Non-Target & 17.74 & 15.49 \\
Both & \bf{19.76} & \bf{18.84} \\
Baseline (DPRNN) & 17.82 & 17.40 \\
\bottomrule
\end{tabular}
\vspace{-15pt}
\end{table}

\section{Conclusions and Future Work}
In this paper, we propose a filter-based model called Dual-Path Filter Network (DPFN) for speech separation. DPFN is the first separation model that focuses on the post-processing of speech separation. It not only takes the advantage of the dual-path structure, but also makes use of the speaker information in the initially separated speech. By providing suitable speaker information for the separation module, DPFN achieves better performance than the baseline DPRNN model. Most importantly, DPFN doesn't need to use PIT in the training phase, and the computational cost is lower. For future work, we want to study more how speaker information guides the speech separation model. We also want to explore more advantages of getting rid of PIT, and consider more types of losses to obtain better speech separation performance.

\newpage
\bibliographystyle{IEEEtran}

\bibliography{references.bib}
\end{document}